\documentclass[twocolumn,showpacs,preprintnumbers,amsmath,amssymb,floatfix]{revtex4}
\usepackage{graphicx}

\begin{document}

\title{H\"older Mean applied to Anderson localization}

\author{Andre M. C. Souza$^{1,2}$, D. O. Maionchi$^{1,3}$  and Hans J. Herrmann$^{3,4}$}

\affiliation{$^{1}$Institut f\"{u}r Computerphysik, Universit\"{a}t Stuttgart, Pfaffenwaldring 27, 70569 Stuttgart, Germany}

\affiliation{$^{2}$Departamento de Fisica, Universidade Federal de Sergipe, 49100-000 Sao Cristovao-SE, Brazil}

\affiliation{$^{3}$Departamento de F\'{i}sica, Universidade Federal do Cear\'{a}, 60451-970 Fortaleza-CE, Brazil}

\affiliation{$^{4}$Computational Physics, IfB, ETH H\"{o}nggerberg, HIF E 12, CH-8093 Z\"{u}rich, Switzerland}

\date{\today}

\begin{abstract}
The phase diagram of correlated, disordered electron systems is calculated within dynamical mean-field theory using the H\"older mean local density of states. A critical disorder strength is determined in the Anderson-Falicov-Kimball model and the arithmetically and the geometrically averages are found to be just particular means used respectively to detect or not the Anderson localization. Correlated metal, Mott insulator and Anderson insulator phases, as well as coexistence and crossover regimes are analyzed in this new perspective. 
\end{abstract}

\pacs{71.10.Fd, 71.27.+a, 71.30.+h}

\maketitle

\section{Introduction}

The properties of materials are strongly influenced by electronic interaction and randomness. In particular, Coulomb correlations and disorder are both driving forces behind metal-insulator transitions (MIT) connected with the localization and delocalization of particles. The Mott-Hubbard MIT is caused by Coulomb correlations (electronic repulsion) in the pure system \cite{Mott}. The Anderson MIT, also referred to as Anderson localization, was established in \cite{cor1} and is the basis of the theory of localization of electrons in disordered systems. It is due to coherent backscattering from randomly distributed impurities in a system without interaction \cite{Anderson}. It is therefore a challenge to investigate quantum models where both aspects are simultaneously present.

The Mott-Hubbard MIT is characterized by opening a gap in the density of states at the Fermi level. At the Anderson localization, the character of the spectrum at the Fermi level changes from a continuous one to a dense discrete one. It is plausible that both MITs could be detected by knowing the local density of states (LDOS), as it discriminates between a metal and an insulator, which is driven by correlations and disorder \cite{pap1}. In many recent works \cite{cor2,cor3,cor4}, the change of the whole LDOS distribution function at the Anderson transition were explored for the Bethe and for the simple cubic lattices, where many other effects, like interaction, percolation and binary alloy disorder, were included.

The theoretical investigation of disordered systems requires the use of probability distribution functions (PDFs) for the random quantities of interest and one is usually interested in typical values of these quantities which are mathematically given by the most probable value of the PDF. In many cases the complete PDF is not known, i.e., only limited information about the system provided by certain averages (moments or cumulants) is available. In this situation it is of great importance to choose the most informative average of a random variable. For examples, when a disordered system is near the Anderson MIT, most of the electronic quantities fluctuate strongly and the corresponding PDFs possess long tails \cite{Mon} and at the Anderson MIT the corresponding moments might not even exist. This is well illustrated by the LDOS of the system. The arithmetic mean of this random one-particle quantity does not resemble its typical value at all. In particular, it is noncritical at the Anderson transition and hence cannot help to detect the localization transition. By contrast, it was shown \cite{pap1,pap2} that the geometric mean gives a better approximation of the averaged value of the LDOS, as it vanishes at a critical strength of the disorder and hence provides an explicit criterion for Anderson localization \cite{Mon,Mon2,pap3,pap4}. Besides that, the work \cite{cor5} suggests that by considering the Anderson localization, the moments of the distribution of the resolvent kernel are of essential importance.

Theoretical descriptions of the MIT have to be nonperturbative if no long-range order exists on either side of the transition. A nonperturbative framework to investigate the Mott-Hubbard MIT in lattice electrons with a local interaction and disorder is given by the dynamical mean-field theory (DMFT) \cite{pap4}. 

In this paper, we investigate the Anderson-Falicov-Kimball model. The pure Falicov-Kimball model describes two species of particles, mobile and immobile, which interact with each other when both are on the same lattice site. The Falicov-Kimball model captures some aspects of the Mott-Hubbard MIT, i.e., upon increasing the interaction the LDOS for mobile particles splits into two subbands opening a correlation gap at the Fermi level. In the Anderson-Falicov-Kimball model the mobile particles are disturbed by a local random potential. The model is solved within the DMFT framework.

Anyway, there is no reason to believe that just the geometric mean can offer a good approximation for the averaged LDOS. Actually, if we consider the generalized mean we can show that the averaged LDOS can vanish in the band center at a critical strength of the disorder for a wide variety of averages. Here, we use the H\"older mean in order to analyze how the averaged LDOS depends on each  H\"older parameter that is used and whether, in particular, the geometric mean can offer a better approximation than the arithmetic one.

A generalized mean, also known as H\"older mean, is an abstraction of the Pythagorean means including arithmetic, geometric and harmonic means. If $q$ is a non-zero real number, we can define the generalized mean with exponent $q$ of the positive real numbers $x_1,\dots,x_n$ as
\begin{equation}
M_q(x)=\left(\frac{1}{n} \sum_{i=1}^n x_{i}^q\right)^{1/q}.
\end{equation}

As special cases, we can find, for example, the minimum ($\lim_{q\to-\infty} M_q(x)$), the geometric mean ($\lim_{q\to0} M_q(x)$), the arithmetic mean ($M_1(x)$) and the maximum ($\lim_{q\to\infty} M_q(x)$). A power mean emphasizes small (big) values for small (big) $q$.

In section \ref{model} we present the  Anderson-Falikov-Kimball model and its solution using the DMFT \cite{pap2}. We also present the relation between the disorder strength and the Coulombian repulsion for the linearized DMFT with $q$-averaging. In section \ref{results} we present the numerical results concerning the ground-state phase diagram obtained using the H\"older mean and compare them with the results of previous works \cite{pap1,pap2,pap3}. Finally in section \ref{conclusions} we present our conclusions and final remarks.

\section{Anderson-Falicov-Kimball Model}\label{model}

\subsection{The Model}

The Falicov-Kimball model is, presently, the simplest model to study
metal-insulator transitions in mixed valence compounds of rare earth and
transition metal oxides, ordering in mixed valence systems, order-disorder
transitions in binary alloys, itinerant magnetism, and crystallization. Recently, it was also applied to study the
possibility of electronic ferroelectricity in mixed-valence compounds, and also of the phase diagram of metal ammonia solutions \cite{Leu-Csa}.
In its most simplified version, namely the static model, it consists in assuming that in the system exist two species of
spinless fermions: one of them possess infinite mass and hence does not move
while the other one is free to move \cite{Andre}. 

In the Anderson-Falicov-Kimball model \cite{pap2} the mobile particles are disturbed by a local random potential, giving rise to a competition between interaction and disorder yielding stabilization of metalicity. This model is defined by the following Hamiltonian:
\begin{equation}
H=t\sum_{<ij>}c_{i}^{+}c_{j}+\sum_{i}\epsilon_{i}c_{i}^{+}c_{i}+U\sum_{i}f_{i}^{+}f_{i}c_{i}^{+}c_{i},\label{Hamil}
\end{equation}
where $c_{i}^{+}$ ($c_{i}$) and $f_{i}^{+}$ ($f_{i}$) are, respectively, the creation (annihilation) operators for the mobile and immobile fermions (electrons and ions, respectively) at a lattice site $i$, $t$ is the electron transfer integral connecting states localized on nearest neighbor sites and $U$ is the Coulombian repulsion that operates when one ion and one electron occupy the same site. The average number of electrons (ions) on site $i$ is denoted as $n_{e}=c_{i}^{+}c_{i}$ ($n_{f}=f_{i}^{+}f_{i}$).The energy $\epsilon_{i}$ is a random, independent variable, describing the local disorder disturbing the motion of electrons. For simplicity, it was assumed that just mobile particles are subjected to the structural disorder \cite{pap2}.

The number of ions and electrons is considered independent of each other and fixed. If there is no long-range order, the position of the ions on a lattice is random, what introduces additional disorder apart from that given by the $\epsilon_{i}$ term in the Hamiltonian (\ref{Hamil}). We consider that the occupation $n_{f}$ on the $i$th site has probability $p$ ($0<p<1$).

In the $U$ term in the Hamiltonian (\ref{Hamil}) one has to take the quantum-mechanical average over a given quantum state of the $f$ particles, what does not transform the extended states into the localized ones. In the $\epsilon_{i}$ term one has to average the quantum-mechanical expectation values over different realizations of $\epsilon_{i}$, what can lead to the Anderson localization \cite{pap2}.

In the pure Falicov-Kimball model, if $n_{e}+n_{f}=1$, the Fermi energy for electrons is inside of the correlation (Mott) gap opened by increasing the interaction \cite{pap2}. How the disorder changes this gap and how the $q$-averaging influences the presence of localized states are the subjects of the present study. 

\subsection{Dynamical mean-field theory}

The Anderson-Falicov-Kimball model, where the interaction and disorder are local, is solved within the DMFT. The formalism that is used is the same from \cite{pap2}, where $\omega$ denotes the energy, the number of the immobile particles is conserved being zero or one and the chemical potential $\mu$ is introduced only for the mobile subsystem. 

Using the hybridization function $\eta(\omega)$, which is a dynamical mean field (molecular field) describing the coupling of a selected lattice site with the rest of the system \cite{pap3}, the $\epsilon_{i}$-dependent LDOS is \cite{pap2}
\begin{equation}
\rho(\omega,\epsilon_{i})=-\frac{1}{\pi}Im G(\omega,\epsilon_{i}),\label{eq1}
\end{equation}
where $G(\omega,\epsilon_{i})$ is the local $\epsilon_{i}$-dependent Green function described in \cite{pap2}.

From the $\epsilon_{i}$-dependent LDOS, we introduce the $q$-averaged LDOS
\begin{equation}
\rho_{q}(\omega)=\left\{\sum_{i} [\rho(\omega,\epsilon_{i})]^{q}\right\}^{1/q}.\label{eq3}
\end{equation}\label{aq}
where the subscript $q$ stands for the chosen average.

The lattice, translationally invariant, Green function is given by the corresponding Hilbert transform
\begin{equation}
G(\omega)= \int d\omega' \frac{\rho_{q}(\omega')}{\omega-\omega'}.\label{eq4}
\end{equation}

\begin{figure}[b]
\includegraphics[width=85 mm,height=60 mm]{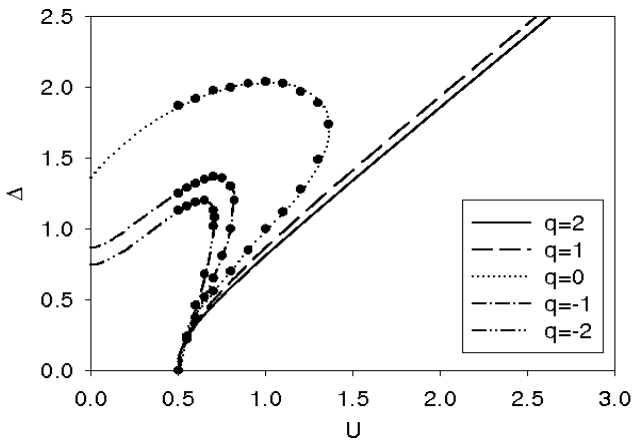}
\caption{Ground-state phase diagram for electrons in a band center determined by using the $q$-average. Dots are determined from the numerical solution of the DMFT equations. Solid lines are obtained by numerical integration from the linearized DMFT.}\label{cur}
\end{figure}

The self-consistent DMFT equations are closed through the Hilbert transform $G(\omega)=\int d\epsilon N_{0}(\epsilon)/[-\epsilon+\eta(\omega)+1/G(\omega)]$, where $N_{0}(\epsilon)$ is the non-interacting density of states.
Using the semi-elliptic density of states for the Bethe lattice, that is $N_{0}(\epsilon)=4\sqrt{1-4(\epsilon/W)^2}/(\pi W)$ \cite{Bethe}, the Anderson-Falicov-Kimball model can be exactly solved. In this case, we can find $\eta(\omega)=W^2 G(\omega)/16$ \cite{pap2}. We considered $\epsilon_{i}$ an independent random variables characterized by a probability function $P(\epsilon_{i})$. We assume here a box model, i.e., $P(\epsilon_{i})=\Phi(\Delta/2-\epsilon_{i})/\Delta$, with $\Phi$ as the step function. The parameter $\Delta$ is a measure for the disorder strength. If we consider the real and imaginary parts of $\eta(\omega)$ we can use $r(\omega)+is(\omega)=\frac{G(\omega)}{16}$, what leads to
\begin{equation}
\rho(\omega,\epsilon_{i})=-\frac{s}{\pi}\frac{\alpha_{i}^2+s^2+(U/2)^2}{[\alpha_{i}^2+s^2+(U/2)^2]^2-U^2\alpha_{i}^2}\label{eq2}
\end{equation}
where $\alpha_{i}=\omega -\epsilon_{i}-r$.

The chemical potential $\mu=U/2$, corresponding to a half-filled band (i.e., $n_{e}=1/2$), and $p=1/2$ are assumed in this paper. $W=1$ sets the energy units.

\subsection{Linearized Dynamical Mean-Field Theory}

At the MIT, the LDOS vanishes in the band center and is arbitrarily small in the vicinity of the MIT (but on the metallic side). The ground-state properties in the half-filled band case are solely determined by the quantum states in the band center ($\omega=0$). In this sense, by linearizing the DMFT equations we can determine the transition points on the phase diagram. Due to the symmetry of $\rho_{q}(\omega)$, in the band center we have that $G(0)=-i\pi \rho_{q}(\omega)$ and is purely imaginary \cite{pap2}, what leads to the recursive relation $\eta^{(n+1)}(0)=-i\pi\rho_{q}^{n}(0)/16$. The left hand side in the $(n+1)$th iteration step is given by the result from the $(n)$th iteration step.  Using Eq. (\ref{eq2}) and expanding it with respect to small $\rho_{q}^{n}(0)$, we find  that the recursive relations within the linearized DMFT with $q$-averaging are
\begin{equation}
\rho_{q}^{n+1}(0)=\frac{1}{16}\rho_{q}^{n}(0)\left[\frac{1}{\Delta}\int_{-\Delta/2}^{\Delta/2} Y(\epsilon)^q d\epsilon\right]^{1/q},
\end{equation}
where
\begin{equation}
Y(\epsilon)=\frac{\epsilon^2+(\frac{U}{2})^2}{[\epsilon^2-(\frac{U}{2})^2]^2}.
\end{equation}

At the boundary curves between metallic and insulating solutions the recursions are constant, $\rho_{q}^{n+1}(0)=\rho_{q}^{n}(0)$, because in a metallic phase the recursions are increasing, whereas in the insulating phase they are decreasing. This observation leads directly to the exact (within DMFT) equations determining the curves $\Delta_{q}=\Delta_{q}(U)$, i.e.,
\begin{equation}
\Delta_{q}=\left(\frac{1}{16}\right)^q\int_{-\Delta_{q}/2}^{\Delta_{q}/2} Y(\epsilon)^q d\epsilon\label{du}
\end{equation}
for the linearized DMFT with $q$-averaging.

\begin{figure}[b]
\includegraphics[width=85 mm,height=60 mm]{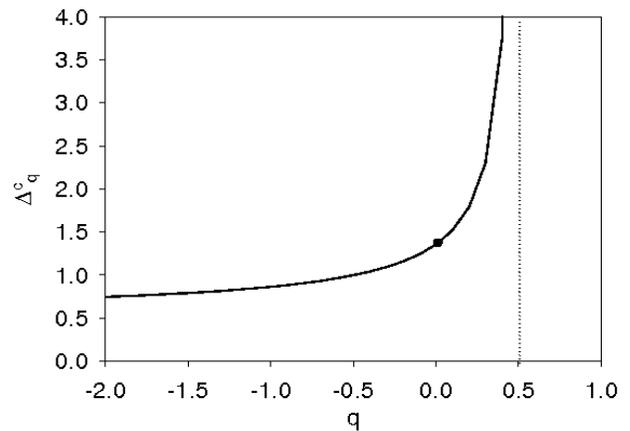}
\caption{$\Delta_{q}^c$ as a function of $q$ for $U=0$. The point represents the exact result for the cubic lattice corresponding to $q=0.011424$ and $\Delta_{q}^c=1.375$.}\label{deltaq}
\end{figure}

\section{Results}\label{results}

\subsection{The critical disorder strength}

In the limit $U=0$, the equation (\ref{du}) can be analytically solved and we can find the critical $\Delta$ as a function of $q$
\begin{equation}
\Delta_{q}^c(0)=\frac{1}{2}\left(\frac{1}{1-2q}\right)^{1/2q}
\end{equation}

We plot this function in Fig. \ref{deltaq}. As $\Delta_{q}^c(0)$ is real, finite and positive, we must have $q<1/2$.

This shows that the Anderson localizations can be detected when we consider a generalized mean with $q<1/2$. In this context, one can easily note that the geometrical and the arithmetical averages ($q=0$ and $q=1$, respectively) are just particular means, where, respectively, the Anderson localization can either be detected or not.
\begin{figure}[b]
\includegraphics[width=80 mm,height=120 mm]{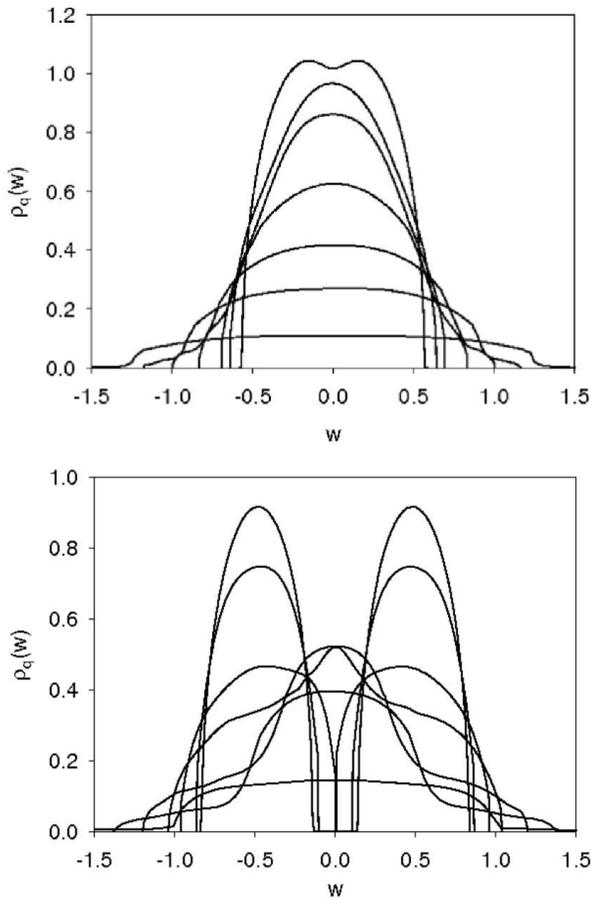}
\caption{Averaged local density of states at $q=0.5$ for $U=0.3$ (upper panel) and $U=0.9$ (lower panel) for different disorder strengths $\Delta$ ($0,0.4,0.8,1.2,1.6,2.0$ and $2.4$).}\label{graf1}
\end{figure}

The numerical integration of equation (\ref{du}) is shown as solid curves in Fig. \ref{cur} for different $q$'s. We can clearly see that for values of $q$ greater than $0.5$, $\Delta$ always increases with $U$, what characterizes the Mott-Hubbard MIT. For values of $q$ smaller than $0.5$ we can always identify three regions with respect to the states at the Fermi level: extended gapless phase, localized gapless phase and gap phase \cite{pap2}. 

As already pointed out in \cite{pap2}, depending on the spectral properties of the Anderson-Falicov-Kimball model, we distinguish three different regimes: weak, intermediate  and strong interaction regime. In the weak interaction regime the Mott gap is not open; in the intermediate interaction regime, the Mott gap is open at $\Delta=0$ and in the strong interaction regime the Mott gap, determined within the DMFT framework with $q$ less than $0.5$, is always open even at large disorder.

The Mott and Anderson insulators are continuously connected. Hence, by changing $U$ and $\Delta$ it is possible to move from one type of insulator to the other without crossing the metallic phase. This occurs because the Anderson MIT($U=0$) and the Mott-Hubbard MIT ($\Delta=0$) are not associated with a symmetry breaking \cite{pap2}.

When we look at the curves obtained for $q$ smaller than $0.5$, where the Anderson localization can be detected, we see (Fig. \ref{cur}) that the curves obtained for greater values of $q$ enclose those ones for smaller values of $q$. If we look at the curve corresponding to the smallest value of $q$ we see that in order to obtain the Anderson localization, one needs to consider a small range of values of $U$ and $\Delta$; in the other extreme, for values of $q$ close to $0.5$ the range of values of $U$ that can be considered is large, but one needs extremely high values of $\Delta$ to detect it.

\subsection{The LDOS}

\begin{figure}[t]
\includegraphics[width=85 mm,height=60 mm]{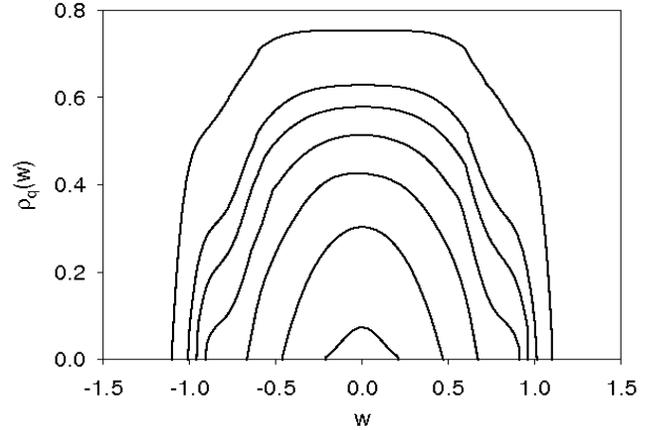}
\caption{Averaged local density of states at $U=0.3$, for $\Delta=1.4$ and $q=-0.25$, $0$, $0.25$, $0.5$, $0.75$, $1.0$ and $2.0$. As $q$ decreases $A_{q}(0)$ is smaller.}\label{var1}
\end{figure}

In order to understand how the numerical solutions of the curves of Fig. \ref{cur} are obtained one needs to observe at first the behavior of the average local density of states for different values of $U$ as the disorder strength $\Delta$ is varied. This is made considering equations (\ref{eq1})-(\ref{eq4}), from which we can obtain the stable value of $\rho_{q}(\omega)$ as a function of $\omega$, as can be seen in Fig. \ref{graf1}. 
\begin{figure}[t]
\includegraphics[width=80 mm,height=180 mm]{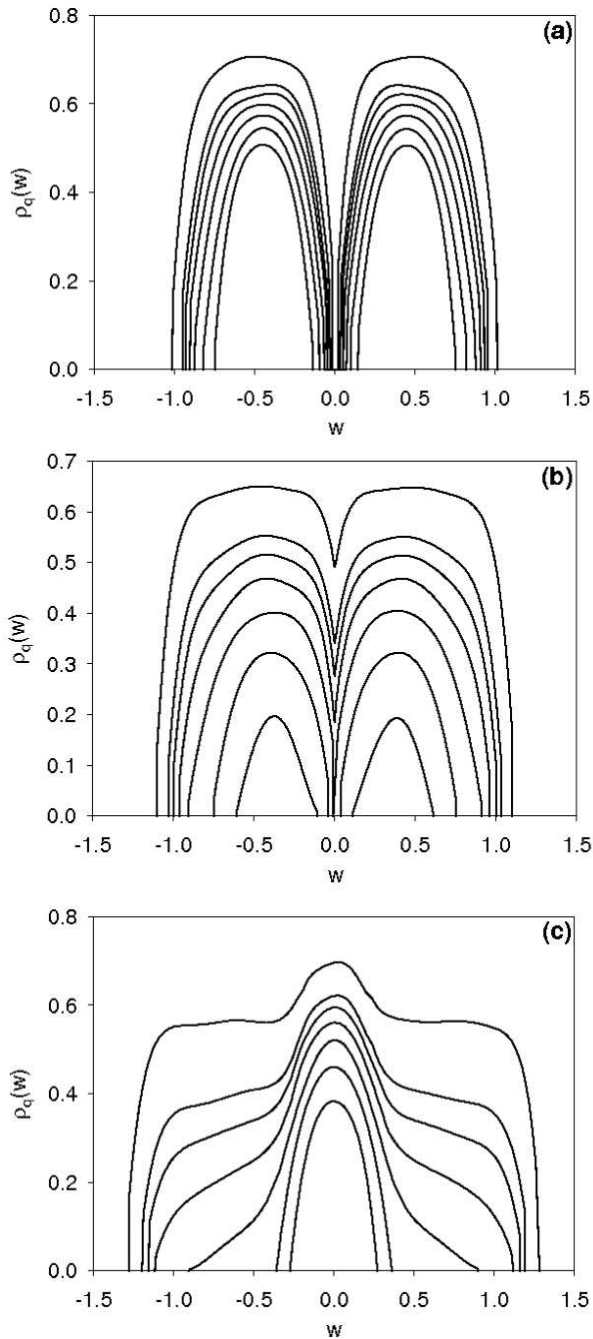}
\caption{Averaged local density of states at $U=0.9$ for $q=-0.25$, $0$, $0.25$, $0.5$, $0.75$, $1.0$ e $2.0$ and for (a) $\Delta=0.6$, (b) $0.8$ and (c) $1.2$. As $q$ decreases $A_{q}(0)$ is smaller.}\label{var2}
\end{figure} 

\begin{figure}[b]
\includegraphics[width=110 mm,height=80 mm,angle=-90]{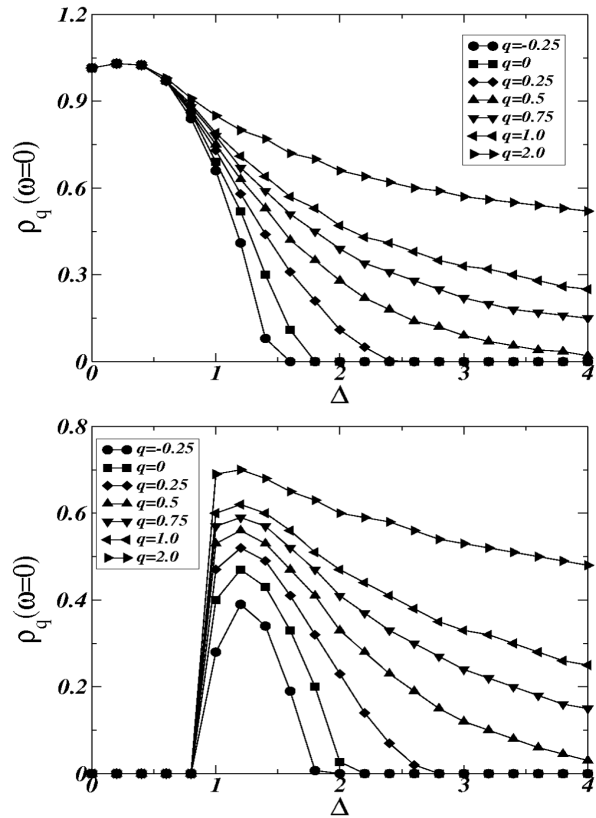}
\caption{Average density of states in a band center ($\omega=0$) as a function of disorder $\Delta$ with $U=0.3$ (upper panel) and $U=0.9$ (lower panel). The results are obtained for $q=-0.25$, $0$, $0.25$, $0.5$, $0.75$, $1.0$ e $2.0$.}\label{ult}
\end{figure}

In all simulations we considered that the initial value of $\rho(\omega,\epsilon_{i})$ is a uniform distribution with bandwidth $W=2 t$ and then we determined $G(\omega)$ in order to obtain $\eta(\omega)$ and finally the new values of $\rho(\omega,\epsilon_{i})$. This loop is performed until we find the stable configuration for $\rho_{q}(\omega)$.

If we consider fixed values of $U$ and $\Delta$ we can observe the dependence of the averaged LDOS on the value of $q$ that is used. Some of these results are presented in Fig. \ref{var1} and Fig. \ref{var2}. In the first one, the simulations are performed in a weak interaction regime ($U=0.3$). In the second one, we consider an intermediate interaction regime ($U=0.9$). Although the behavior of the LDOS is not the same in both cases, we can see that the smaller the value of $q$ that we consider, the smaller is the value of $\rho_{q}(\omega=0)$. We can also see that the shape of the curves change when we compare the curves for $q$ smaller than $0.5$ and greater than $0.5$.

A signature of the Anderson localization is the vanishing of $\rho_{q}(\omega=0)$ as we increase $\Delta$. The way to obtain the results in Fig. \ref{cur} consists in considering a range of values of $U$ for a fixed $q$ and varying the disorder strength $\Delta$ for each value of $U$, in order to determine the values of $\Delta$ when $\rho_{q}(\omega=0)=0$. 

In the simulations, we used values of $\Delta$ varying of $0.2$ and determined the stable values of $\rho_{q}(\omega=0)$ for each of them. As we use an iterative process, the values of $\rho_{q}(\omega=0)$ that we obtain converge always to the stable one. In this sense, for some values of $\Delta$ the stable value of $\rho_{q}(\omega=0)$ is exactly null. In Fig. \ref{ult} we show the dependence of $\rho_{q}(\omega=0)$ on $\Delta$ for two different values of $U$ ($0.3$ and $0.9$) that, again refers to the weak and intermediate interaction regime, respectively. For the first one, increasing $\Delta$ at fixed $U$ and $q$ leads to further decreasing of the LDOS; just for values of $q$ smaller than $0.5$, $\rho_{q}(\omega=0)$ can vanish for a finite value of $\Delta$; for $q$ greater than $0.5$ this would just happen at infinity. For $q$ smaller than $0.5$ we found that the critical exponent of the curves $\rho_{q}(\omega=0) \approx (\Delta-\Delta_{c})^\beta$ is always $\beta=1$ \cite{pap3}, which guarantees that they all belong to the same universality class. For the second plot, $\rho_{q}(\omega=0)$ is zero until a certain value is reached, which corresponds to the transition from the gap phase to the extended gapless phase. Here again just for values of $q$ smaller than $0.5$, $\rho_{q}(\omega=0)$ turns to vanish for a finite  value of $\Delta$. Performing the same process for other values of $U$ we obtain easily all the simulation points of Fig. \ref{cur} for a variety of different values of $q$.

\section{Conclusions}\label{conclusions}

In the present paper, we generalized the solutions of the Anderson-Falicov-Kimball model using a H\"older mean for the averaged local density of states. In this context, where each mean is characterized by the H\"older parameter $q$, we found that the averaged LDOS can be calculated for arbitrary parameters in order to detect the Anderson-localization. As we vary the parameter in the H\"older mean, we emphasize different signal-values of the LDOS.

Besides that, there is a critical value for this parameter, that is $q=1/2$. For values of $q$ greater and smaller than this, the dependence of the critical strength of disorder $\Delta$ on the Coulombian repulsion $U$ is, respectively, similar to the one found in \cite{pap2} using the arithmetic mean ($q=1$) and the geometric mean ($q=0$). We showed that not only the geometric mean can be used to detect the Anderson mean, but also an infinity of other averages.  One example occurs for the cubic lattice \cite{pap3,rd}, where the exact solution $\Delta_{q}^c=1.375$ corresponds to $q=0.011424$ in our model (Fig. \ref{deltaq}), which is not exactly the geometric mean. 

In this way, using the H\"older mean, we were able to understand why in many previous works the geometric mean has offered a better approximation for the averaged LDOS than the arithmetic one.

\section{Acknowledgments}
The work was supported by Conselho Nacional de Pesquisas Cientificas (CNPQ) and Deutsche Akademische Austauschdienst (DAAD). H.~J~Herrmann thanks the Max-Planck prize.

\end{document}